\documentclass[journal,twoside,web]{ieeecolor}
\usepackage{generic}
\usepackage{cite}
\usepackage{amsmath,amssymb,amsfonts}
\usepackage{algorithmic}
\usepackage{graphicx}
\usepackage{algorithm,algorithmic}
 
\usepackage{booktabs}
 
\usepackage{multirow}

\usepackage{hyperref}
\usepackage{textcomp}
\def\BibTeX{{\rm B\kern-.05em{\sc i\kern-.025em b}\kern-.08em
    T\kern-.1667em\lower.7ex\hbox{E}\kern-.125emX}}
\begin{document}

\title{Dynamic Brain Behaviours in Stroke: A Longitudinal Investigation Based on fMRI Analysis}

\author{Kaichao Wu, Beth Jelfs, Katrina Neville, and Qiang Fang* 
\thanks{This work was supported by the Li Ka Shing Foundation Cross-Disciplinary Research Grant (2020LKSFG01C). The asterisk indicates the corresponding author.}
\thanks{Kaichao Wu, and *Qiang Fang are with the Department of Biomedical Engineering, College of Engineering, Shantou University, Shantou, China (e-mail: \{kaichaowu, qiangfang\}@stu.edu.cn). }
\thanks{Kaichao Wu and Katrina Neville are with the School of Engineering, RMIT University, Melbourne, Australia (e-mail: katrina.neville@rmit.edu.au).}
\thanks{Beth Jelfs is with the Department of Electronic, Electrical, and Systems Engineering, University of Birmingham, Birmingham, UK. (e-mail: b.jelfs@bham.ac.uk.) }
}
\maketitle

\begin{abstract}
Background: The brain's functional network constantly adapts to external changes. However, the mechanisms underlying this dynamic adaptive behavior in stroke patients with motor injuries and its role in post-stroke motor recovery remain poorly understood.

Method: This study conducted a long-term investigation involving 15 first-stroke patients. Each participant underwent five fMRI scans distributed equally over a six-month period. Using functional neuroimaging data, time-varying functional modularity in post-stroke patients was detected, and subsequently, the dynamic brain behaviors, including recruitment, integration, and flexibility, along with their longitudinal changes, were assessed.

Results: Our findings reveal that stroke lesions lead to significant and enduring alterations in all three dynamic behaviors within functional brain networks. Furthermore, during the six-month recovery period, patients who exhibited good and poor recovery showed notable differences in recruitment and flexibility, indicating distinct recovery trajectories for these groups. Notably, when predicting post-stroke recovery status, whole-brain recruitment emerged as a robust and reliable feature, achieving an AUC of 85.93

Significance: Our study offers a comprehensive depiction of dynamic brain behavior in the post-ischemic-stroke brain, with a focus on longitudinal changes concurrent with functional recovery. These dynamic patterns hold promise as valuable tools for evaluating and predicting motor recovery following stroke

\end{abstract}

\begin{IEEEkeywords}
Dynamic, Brain behavior, Network reorganization, motor recovery, Stroke.
\end{IEEEkeywords}

\section{Introduction}
\label{sec:introduction}

\IEEEPARstart{S}{troke} is the principal cause of adult long-term disability throughout the globe ~\cite{RN1}, leaving a majority of stroke survivors to suffer from motor impairment ~\cite{RN2, RN3}. The mechanisms underpinning motor function recovery after stroke are intricately tied to adaptive changes within the brain's functional networks driven by neuroplasticity ~\cite{RN4, RN6, RN5}. Consequently, establishing a comprehensive understanding of brain adaptive behaviors following stroke is of prime importance to successfully design effective rehabilitation strategies and interventions. In this context, magnetic resonance imaging (MRI), particularly functional MRI (fMRI), has become one of the most prominent ways of explicating brain adjustment behind post-stroke motor recovery ~\cite{RN34, RN33}. The most common brain behavior, in this regard, is the reinforced functional network connectivity (FC)  between hemispheres (e.g., the ipsilesional primary motor cortex (M1) and contralesional M1 ~\cite{RN9, RN8, RN15}), as well as between motor-relevant regions (e.g., SMA~\cite{RN32} and other motion-related regions like the dorsolateral prefrontal cortex (PMD)~\cite{RN31, RN17}). These enhancements in regional FC have consistently shown significant correlations with improvements in motor function. In parallel with these changes, other adaptive behaviors within the brain's network dynamics have also been identified. These include the regeneration of FC between regions affected by stroke-induced lesions ~\cite{RN16, RN15}, as well as the normalization of brain modularity ~\cite{RN18}. Both of these adaptive behaviors have demonstrated concurrent associations with functional recovery.

These adjustment behaviors can be summarized with an umbrella term of network plasticity or network reorganization~\cite{RN35, RN38, RN37}. However, it's crucial to note that many of these findings primarily rely on static FC measures, which offer insights based on long-term averages~\cite{RN36}. While these static methods provide valuable information, they may fall short of capturing the inherent dynamics of the brain, particularly in the context of post-stroke recovery, which is a time-dependent process. In contrast, recently developed techniques in the form of time-varying functional network connectivity (DFNC) analyses have emerged, allowing for the examination of moment-to-moment fluctuations in FC strength ~\cite{RN22, RN23}. With DFNC, researchers can analyze brain adaptive behaviors on a timescale of seconds, linking their temporal dynamics directly to the process of post-stroke recovery. 

\begin{figure*}[!t]
    \centering
    \includegraphics[width = \textwidth]{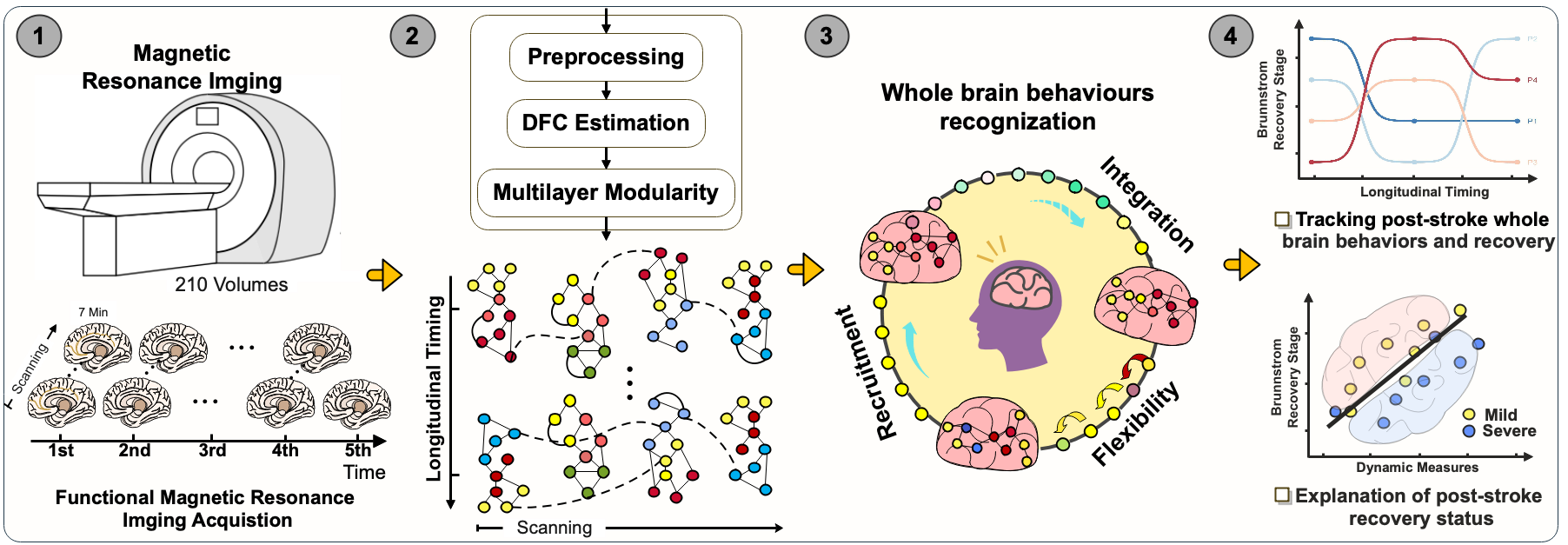}
    \caption{The intact analytical pipeline of such longitudinal investigation. (1) Longitudinal clinical and neuroimaging data collection; (2) Dynamic evaluation of functional modularity. (3) Whole-brain behavior recognition, and (4) Post-stroke behavior and recovery analysis.\label{fig: Pipeline}}
\end{figure*}

The dynamic brain adaptive behaviors provide a novel insight into unscrambling the post-stroke motor function recovery. Whereas, like many other complex systems, the flexibility and adaptability of the brain are supported by its modular structure ~\cite{RN24}, which suggests dynamic behaviors stemming from whole brain function modules should be explored. Intuitively, brain function modules are a group of functional regions highly connected internally but weakly linked externally ~\cite{RN28}. The function modularity allows the brain to switch between segregated and integrated states in order to meet the motor demands ~\cite{RN30, RN29} or to respond to the altered external task ~\cite{RN27}. The modularity alterations, particularly its time-varying patterns, have contributed to understanding the role of brain adaptability in a variety of neurological diseases (schizophrenia ~\cite{RN39}, temporal lobe epilepsy ~\cite{RN40} and depression ~\cite{RN41}). However, how the post-stroke brain dynamically adjusts its modular structure after stroke and how this behavior supports post-stroke motor recovery has never been thoroughly investigated.

In this work, a cohort of 15 stroke patients experiencing motor impairment was recruited, Over a span of 6 months, five fMRI scans for each participant were collected. These functional neuroimages were used to investigate the brain adaptive behaviors (recruitment, integration, flexibility) derived from dynamic function modularity. There are three aims in this study. First, we wish to identify whole brain behaviors that can reflect the inherent dynamic function modularity and determine whether the stroke attacks induce alteration. Secondly, we delve into the changes in brain behaviors over the course of the motor recovery period following a stroke and examine the difference between patients with distinct motor outcomes. Finally, inspired by the previous study on the relevance of FC dynamics and post-stroke function outcomes~\cite{RN19, RN20, RN21}, we used the machine learning method to explain the recovery status of motor function with the help of whole-brain behaviors, to enhance the understanding of post-stroke neural reorganization and recovery mechanisms.

\section{METHOD AND MATERIAL}
\subsection{Participant}
The stroke samples examined in this study were from fifteen ischemic stroke patients(ISP) admitted to the First affiliated hospital of Shantou University Medical College (6 right and 9 left ISP, mean age is 63.8 with standard deviation 11.68 years, 4 males, mean day of first MRI scan post stroke 23.06 with standard deviation 4.32). The patients were recruited from a study that has been approved by the medical research ethics committees of the named hospitals, and all participants signed informed consent. In addition, fifteen age-matched healthy samples were served as control groups (7 males, the mean age is 68.6 with a standard deviation of 6.4 years).

\begin{table*}[tbp]
\centering
    \fontsize{28pt}{28.5pt}\selectfont
    \caption{Demographics and clinical characteristics of participants.\label{tab1}}
    \vspace{6pt}
    \renewcommand\arraystretch{1.5}
    \resizebox{\textwidth}{!}{
        \begin{tabular}{lcccccccccccc}
            \toprule[3pt]
            \multirow{2}{*}{Patient} & \multirow{2}{*}{Sex} & \multirow{2}{*}{Age} & \multirow{2}{*}{\begin{tabular}[c]{@{}l@{}}~~Side\\of Lesion\end{tabular}} & \multirow{2}{*}{Lesion location}&  \multirow{2}{*}{\begin{tabular}[c]{@{}l@{}}Lesion Volumes\\\hspace{2cm}($CM^3$)\end{tabular}} & \multirow{2}{*}{\begin{tabular}[c]{@{}l@{}}Day Since \\    \hspace{1cm}Stroke\end{tabular}} &  \multicolumn{5}{l}{\hspace{9cm}Timing BS } & \multirow{2}{*}{Recovery} \\ 
            \cline{8-12}
            &   &  &    &   &  &  & 1st   & 2nd   & 3rd   & 4th   & 5th  &   \\      
           \midrule[3pt]
            \multicolumn{13}{l}{ \textit{\textbf{Good recovery group}}} \\
            P2  & F & 65 & R & Cerebellum & 0.91  & 26 & 2 & 4 & 5 & 6 & 6 & good \\
            P3  & M & 54 & L & Thalamus  & 3.61  & 42 & 1 & 2 & 3 & 3 & 3 & good \\
            P4  & M & 58 & L & Temporal lobe/basal ganglia & 1.53  & 22 & 4 & 5 & 6 & 6 & 6 & good \\
            P11 & F & 78 & R & Posterior limb of internal   capsule & 2.39  & 22 & 1 & 1 & 3 & 3 & 3 & good \\
            P13 & M & 47 & L & Precentral/   Mid frontal            & 6.23  & 18 & 3 & 4 & 5 & 5 & 5 & good \\
            P14 & M & 57 & L & Pallidum                             & 6.44  & 18 & 1 & 3 & 3 & 4 & 4 & good \\
            
            \textit{Mean$\pm$SD}  & - & \textit{59.42$\pm$9.05}  & - & -                            & \textit{3.93$\pm$2.24}   & \textit{24.66$\pm$8.21}  & \textit{2$\pm$1.06}  & \textit{3.16$\pm$1.34}  & \textit{4.16$\pm$1.21}  & \textit{4.50$\pm$1.25}  & \textit{4.50$\pm$1.25}  & good \\
            
            \multicolumn{13}{l}{\textit{\textbf{Poor recovery group}}} \\
            P1  & M & 59 & R & Temporal/occipit lobe                & 63.55 & 16 & 5 & 5 & 5 & 6 & 6 & poor \\
            P5  & M & 63 & R & Thalamus/Hippcocampus                & 8.68  & 20 & 3 & 2 & 2 & 3 & 4 & poor \\
            P6  & F & 67 & R & Thalamus/cuneus/lingual              & 39.81 & 23 & 1 & 1 & 1 & 1 & 1 & poor \\
            P7  & F & 48 & L & Thalamus                             & 0.95  & 21 & 6 & 6 & 6 & 6 & 6 & poor \\
            P8  & M & 48 & L & Pallidum                             & 1.09  & 24 & 6 & 6 & 6 & 6 & 6 & poor \\
            P9  & M & 80 & L & Mid Occipital/lingual                & 50.51 & 26 & 1 & 1 & 2 & 2 & 2 & poor \\
            P10 & M & 78 & L & Thalamus                             & 3.12  & 27 & 6 & 6 & 6 & 6 & 6 & poor \\
            P12 & M & 81 & L & Hippocampus                          & 1.04  & 27 & 1 & 1 & 1 & 1 & 1 & poor \\
            P15 & M & 73 & L & Putamen                              & 5.33  & 14 & 6 & 6 & 6 & 6 & 6 & poor  \\
            
            \textit{Mean$\pm$SD} & - & \textit{66.33$\pm$12.11}  & - & -                            & \textit{19.43$\pm$19.54}   & \textit{22$\pm$4.42}  & \textit{3.88$\pm$2.23}  & \textit{3.77$\pm$2.29}  & \textit{3.88$\pm$2.21}  & \textit{4.11$\pm$2.18 } & \textit{4.11$\pm$2.28}  & poor \\
            \bottomrule[3pt]
\multicolumn{13}{l}{ \textit{Abbreviations: R: Right, L: Left, BS: Brunnstrom stage, SD: Standard deviation. 1st: 20-30 days , 2nd: 50-60 days, 3rd: 80-90 days, 4th: 110-120 days. 5th: 140-150 days.} } 
        \end{tabular}
}
\end{table*}

\subsection{Motor performance measurement}
The motor recovery of stroke patients was quantified by the Brunnstrom stage method \cite{shah1986stroke} with an interval of 30-40 days in 6 six months after stroke. This method is widely used in stroke research and assesses which recovery stage the patients were at according to their upper limb function in four dimensions (i.e. grasp, grip, pinch, and gross movements; range 1–6; 1 = unable to perform any movements, 6 = normal movement). The recovery process was assessed in front of the MRI room before patients completed the scan. Patients whose Brunnstrom stage improved by 2 stages were considered to have good recovery, while those whose Brunnstrom stage had narrow or minor improvement were considered to have poor recovery. The demographic characteristics of all participants and the clinical features of stroke patients can be seen in \autoref{tab1}. To assess their severity, patients with Brunnstrom stage $<$ 3 were assigned to a severe attack subgroup, otherwise, they would belong to a mild attack group.

\subsection{The analytical pipeline}

The analytical pipeline in this study comprises four key components, as illustrated in \autoref{fig: Pipeline}: (1) Longitudinal clinical and neuroimaging data collection; (2) Dynamic evaluation of functional modularity. (3) Whole-brain behavior recognition, and (4) Post-stroke behavior and recovery analysis. The technical details pertaining to each part are described in the following sections.

\subsection{The longitudinal clinical and neuroimaging data collection}

Resting-state functional MRI scans ( 3.0T, Discovery) were collected once equally 30 to 40 days within six months. The high-resolution T1 anatomical images were acquired for structure reference, with 1 mm isotropic voxels, a 256 × 256 matrix size, and a 9-degree flip angle (129 slices, TR = 2250 ms, TE = 4.52 ms). The fMRI parameters: repetition time (TR) = 2,000 ms; echo time = 30 ms; flip angle = 90; field of view = 240 *240 mm2; matrix size = 64 * 64; number of slices = 25; and voxel size = 3.43 *3.43 * 5.0 mm3 with no gap; and 210 volumes acquired in 7 min. 

\subsection{Dynamic evaluation of the brain functional modularity}

The Dynamic evaluation of the brain functional modularity pipeline encompasses three main stages: fMRI data processing, dynamic functional connectivity (DFC) estimation, and multilayer modularity construction. 

\subsubsection{fMRI data processing and denoising}
The functional MRI was processed using a customized preprocessing pipeline in the CONN functional connectivity toolbox. The first ten volumes were discarded from patient scans and control scans respectively to allow a steady blood oxygenation level-dependent activity signal. The remaining 200 functional volumes were continued preprocessed, steps are summarized as follows. (1) realignment for motion correction. (2) slice-time correction. (3) outlier identification using custom artifact detection software that detects outlier time points for each participant, (4) normalization to MNI space and resampling to 3 mm isotropic voxels. 

\begin{figure}[!t]
    \centering
    \includegraphics[width = \columnwidth]{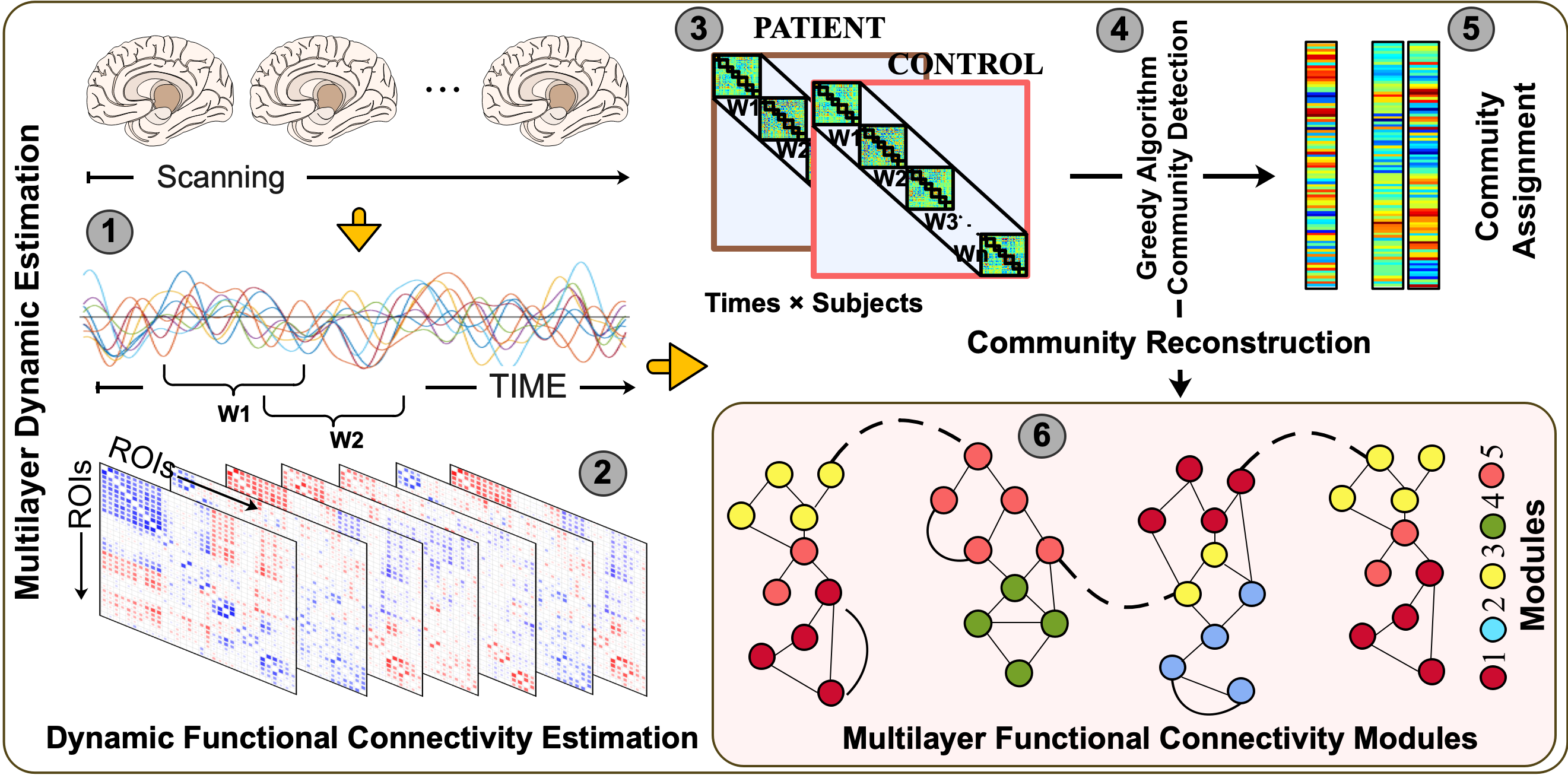}
    \caption{The dynamic functional connectivity estimation and multilayer function connectivity modules detection.\label{fig: multilayer}}
\end{figure}


\subsubsection{Dynamic functional connectivity estimation}
Functional connectivity is a measure of the statistical relation between time series of spatially distinct brain regions. We used the brain parcellation from CONN to extract the whole brain time series from the denoised functional neuroimages (\autoref{fig: multilayer}1). This brain parcellation which is composed of 32 brain regions and 8 function networks is based on recognized ICN from group ICA method and has been used in many studies \textbf{(See Supplementary information for the spatial map}). Then, we calculated the dynamic functional connectivity (DFC) matrixes between the time course (200 time points) with the sliding window scheme (see \autoref{fig: multilayer}2). The window was obtained by convolving a rectangle (equal to the window size) with a Gaussian ($\sigma$ = 3) at each time point. The width is 22 TR (44s) and the shifting step is 1 TR (2.2 s) in this paper, resulting in 179 windows and two adjacent windows having 21 TR overlap. Within each window-length period, we create an N $\times$ N function connectivity matrix calculating the Pearson’s correlation coefficient between the mean signal timeseries. The obtained array of function connectivity matrixes along the time course represents the DFC network in the brain. Finally, Fisher’s Z-transformation was applied to eliminate the bias, with only the positive values being retained in the subsequent analysis.

\subsubsection{Multilayer functional modularity detection}
Following the DFC estimation, the DFC matrixes were concatenated along the diagonal to produce their initial community profile (obtaining a matrix with 5728$\times$5728, 5728 = 179$\times$32 which being the number of sliding windows and the size of each one respectively, see \autoref{fig: multilayer}2). In the next step, the Louvain-like greedy community detection algorithm was used for dynamic community detection. This algorithm optimizes the multilayer modularity partition through maximizing the modularity quality function:
\begin{equation}
    Q_M=\frac{1}{2\mu}\sum_{ijlr}{\left[\left(A_{ijl}-\gamma \frac{k_{il}k_{jl}}{2m_l} \right) \delta _{lr}+\delta_{ij}\omega _{jlr}\right] \delta \left(g_{il},g_{jr}\right)},
\end{equation} 
where \begin{itemize}
    \item $A_{ijl}$ is the weight of the edges between nodes $i$ and $j$ at layer $l$;
    \item $k_{jl}$ is the weighted degree of node $j$ in layer $l$, that is the sum of the weights of the edges connected to node $j$ in layer $l$;
    \item $m_l$ is the total nodal weighted degrees in layer $l$;
    \item $\mu = \frac{1}{2}\sum_{ij}{(k_{jr}+c_{jr})}$ is the sum of the weights of the dynamic functional connectivity matrix;
    \item $c_{jr}=\sum_{l}{\omega_{jrl}}$ and $\omega_{jrl}$ is the edge strength between node $j$ in layer $l$ and node $j$ in layer $r$. 
    \item $\delta_{ij}$ denotes the Kronecker $\delta $-function, where $\delta _{ij}=1$ if $i=j$, otherwise $0$;
    \item $g_{il}$ and $g_{jr}$ represent the community node $i$ is assigned to in layer $l$ and node $j$ in layer $r$ respectively;
    \item $\delta \left( g_{il},g_{jr} \right)=1$ if $g_{ir}=g_{jl}$, otherwise $0$;
    \item The parameters $\gamma$ and $\omega$ are the intra-layer and inter-layer coupling parameters, controlling the number of modules detected in layers and across layers.
\end{itemize}

\subsubsection{ Whole brain behaviors recognition}
 Three brain behaviors were identified based on the detected multilayer functional modularity: recruitment, integration, and flexibility.
 
 The recruitment of a given predefined functional system $S$ is defined as: 
\begin{equation}
    R_S=\frac{1}{n_S}\sum_{i\in S}{\sum_{j\in S}{P_{ij}}},
\end{equation}		
where $n_S$ is the number of ROIs belonging to the system $S$; $P_{ij}$ is the allegiance matrix of the multilayer networks, which is defined as $P_{ij}=\frac{1}{T}\sum_{t=1}^T{a_{ij}^{t}}$ with $a_{ij}^{t}=1$ if in layer $t$ nodes $i$ and $j$ are assigned to the same community, and 0 otherwise. Similar to recruitment, the integration of a given predefined functional system $S$ is defined as:
\begin{equation}
     I_S=\frac{1}{N-n_S}\sum_{i\in S}{\sum_{j\notin S}{P_{ij}}}.
\end{equation}

The system of interest is highly functionally integrated when its functional regions are frequently assigned to the same community as other regions. Therefore, to quantify this, an integration coefficient can also be defined between different functional systems~\cite{RNre}. The integration between functional system $S_k$ and $S_l$ is calculated as:
\begin{equation}
    I_{S_kS_l}=\frac{1}{n_{S_k}n_{S_l}}\sum_{i\in S_k}{\sum_{j\in S_l}{P_{ij}}}.
\end{equation}
The higher the between-system integration, the stronger the functional coordination between systems. This study investigated both within-system and between-system integration alterations caused by stroke lesions.

Flexibility characterizes the community stability of a system in multilayer resolution~\cite{RNfe}. The flexibility of a system corresponds to the average number of times that its brain regions change module allegiance. The system $S$ flexibility is defined as:
\begin{equation}
	     F_S=\frac{1}{n_s \times \left(T-1 \right)}\sum_{i\in S}{\sum_{t=1}^T{b_i}},
\end{equation}
where $n_S$ is the number of regions belonging to the system $S$, $T$ is the multilayer resolution, and $b_i=1$ if in the next layer $t+1$ the node $i$ is assigned to a different community.

Noting that random effects in the Louvain-like greedy community detection algorithm exist in multilayer community detection, the multilayer modularity optimization was run 100 times. The mean of the corresponding dynamic measures from the 100 repetitions served as their final values. Besides, a permutation approach~\cite{RN27} was used for the normalization of these dynamic measures. Specifically, a null distribution was created from 1000 randomly permuted multilayer function connectivity matrices. The recruitment, integration, and flexibility are then divided by the mean of the corresponding null distribution to obtain normalized values.

\subsection{Post-stroke behaviors and recovery analysis}
Post-stroke recovery analysis has two-fold (as \autoref{fig: multilayer}.4 shows): one is tracking the time course of post-stroke brain behaviors and motor recovery; another one is exploring the explanation of post-stroke severity with a machine learning method.

\subsubsection{Tracking post-stroke brain behaviors and motor recovery} 
With the obtained whole brain behavior measurement at five-time points, their changes were modeled in each patient using a log function (given that post-stroke recovery is nonlinear):
\begin{equation}
	     Y = m*log(t)+b
\end{equation}
where $t$ is the days since the stroke, and $Y$ is the brain behavior measurement at each time point. The slope $m$ represents the trends of brain behavior responding to stroke attack and intercept $b$ is an error, and they were solved using least squares fit. Then the slope $m$ was compared between good recovery and poor recovery groups to see if there are significant differences in brain behavior in post-stroke motor performance evolution ( pair t-tests, FDR corrected P$<$0.05). 

To investigate the relationship between whole-brain average behaviors and post-stroke recovery, Pearson correlation coefficients for whole-brain average behaviors measurement and the Brunnstrom stage score were calculated. In addition,  Pearson correlation coefficients were also conducted at different post-stroke phases.

\begin{figure*}[!t]
    \centering
    \includegraphics[width = \textwidth]{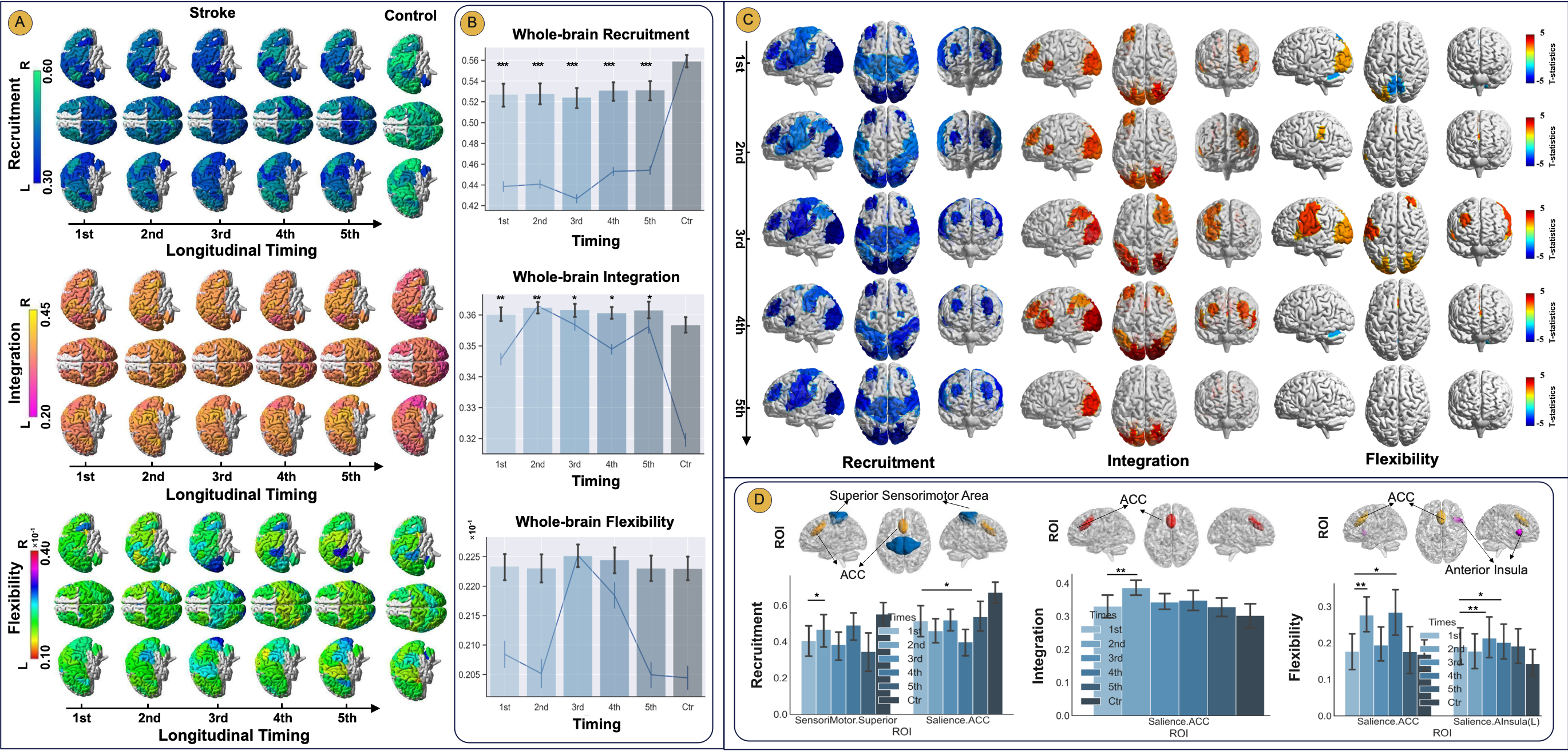}
    \caption{\textbf{A.} The brain map of dynamic brain behaviors across five time points after stroke.\textbf{B.} The trends of the global average in whole brain behaviors over 6 months. \textbf{C.}The brain regions with significantly different whole brain behaviors contrast with healthy controls. \textbf{D.}The brain regions with significant longitudinal change over a period of more than 6 months.   \label{fig: tracking}}
\end{figure*}

\subsubsection{The explanation of post-stroke severity}
At this point, we implemented a ridge regression algorithm (RR) to link whole-brain measurement (independent variables) to the degree of post-stroke motor function (dependent variables). Specifically, all patient’s brain measurements at one-time point will be represented by a regression matrix $X \in R^{N_s \times N_p}$ ( $N_s$ is the number of behaviors measurements, $N_p$ the number of regressors which corresponding to the number of brain regions). This analysis can be used to predict post-stroke motor function, and essentially it can assign a weight $\beta$ to a brain region, indicating its contribution to the post-stroke function assessment. 
Specifically, the training process is to minimize the loss function as below:

\begin{equation}
	     \sum_{i =1}^n{(y_i - \sum_{j=1}^p{x_{ij} \beta_j})^2} + \lambda \sum_{j=1}^p{\beta_j^2},
\end{equation}

and the $\beta$ is estimated as 
\begin{equation}
	\hat{\beta}_{RR} = (X'X+{\lambda}I_p)^{-1}X'Y,
\end{equation}

There are seven RR models (three models for three behavior measurements, three models for the combination of two measurements,s and one for three measurements combined together), and the regularization parameter $\lambda$ was optimized by identifying a value within $[10^{-5}, 10^{5}]$, with 200 logarithmic steps. Specifically, for each value of $\lambda$, each RR model was trained and tested using a leave-one-out-cross-validation loop (LOOCV), which used 75-1= 76 training data to estimate the model weights. The optimal $\lambda$ ($\lambda_{opt}$) value was the one that minimized the prediction error over the training set, and the predictions obtained with $\lambda_{opt}$ were considered as the model regressors.

A receiver operating characteristic (ROC) curve was constructed for each model and the area under the curve (AUC) was calculated to assess the model's accuracy. To obtain the optimal set of RR model weights $\beta$, the weights derived from each LOOCV loop at $\lambda_{opt}$ were averaged across the Ns loops. These selected weights were scaled to [-1, 1] and then back-projected to the brain to display a map of the most predictive brain areas.

\subsection{Statistical analysis }
Our statistical analyses are based on common parametric tests (t-test, f-test). Multiple comparisons correction was always applied whenever testing more than one hypothesis simultaneously (false discovery rate (FDR) correction p $<$ 0.05).

\section{RESULT}

\subsection{The whole brain behaviors and their longitudinal alterations during the post-stroke recovery period}

There were three brain behaviors derived from the dynamic functional modularity (recruitment (integration (Middle), and flexibility (lower)). \autoref{fig: tracking}\textbf{ A }shows the value of these measurements across the whole brain regions,  estimated at five time points in 6 months. 
 It is obviously shown that nearly whole brain recruitment is reduced after stroke (the color bar represents the value range), and this reduction can be observed during the entire six-month recovery period. The alteration of integration, in contrast, is not that distinct, while we can still recognize the increase in the whole brain integration in the six months after stroke.  The flexibility does not show a unifying manner in the whole brain scope. Some regions have dramatic changes, while others do not. The trends of the global average in whole brain behaviors over 6 months can be seen in \autoref{fig: tracking}\textbf{B}. 
 Compared with healthy stroke, significantly reduced recruitment and increased integration can be observed at every month post-stroke. However, no significant changes were observed in flexibility.

\begin{figure*}[!t]
    \centering
    \includegraphics[width = \textwidth]{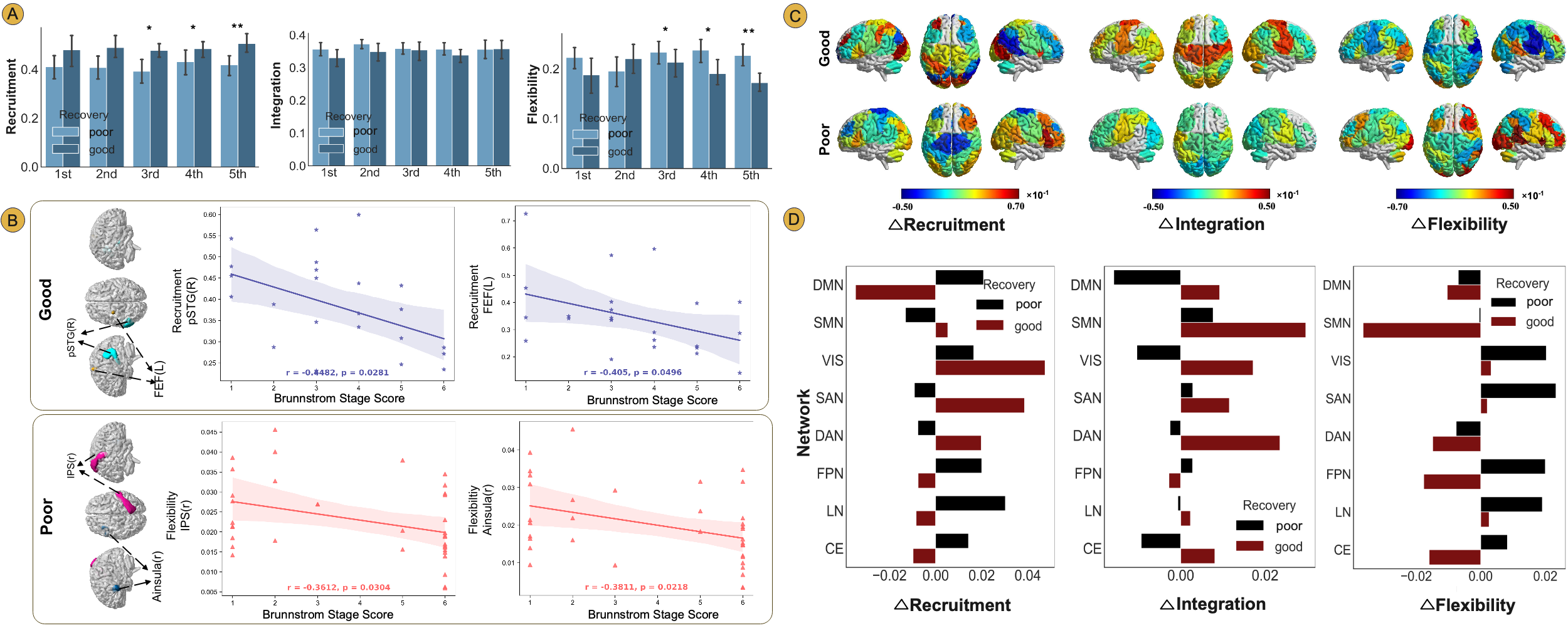}
    \caption{\textbf{A.}Difference in dynamic behavior between good and poor recovery. \textbf{B} significant 
 correlation with brunnstrom stage score. \textbf{C.} the brain map of the recovery of behavior.\textbf{ D.} the recovery of behavior for various networks\label{fig: recovery}}
\end{figure*}

The difference in whole brain recruitment, integration, and flexibility between stroke patients and healthy control at five time points ( Two-sample t-test) were summarized in \autoref{fig: tracking}\textbf{C} and Supplementary material Table I-III. Compared with healthy controls, stroke patients have significantly reduced recruitment and increased integration. The reduced recruitment covered nearly the whole brain, across from the network DAN to VIS, the first month after the stroke. However, the reduced regions decreased with time goes up. Six months later, recruitment can be mainly seen declined in network SAN, SMN, and VIS, compared to healthy controls. The reduced recruitment can always be observed in the SMN network, from both the superior to the lateral sensorimotor area. The significantly increased integration mainly resides in network SAN and VIS. Likewise, a raised integration at all times after stroke but in the network VIS. 

In terms of flexibility, the group effect of flexibility did not show consistency in the network. Significantly lower flexibility can only be observed in the anterior cerebellum at 20-30 and 110-120 days after stroke, and in the PCC at 20-30 days after stroke. In contrast, the regions with significantly increased flexibility can be seen at the most time, from lateral visual and sensorimotor areas to most SAN regions. Of note that at the end of more than 6 months, stroke patients do not show a difference in the whole brain flexibility.

Next, we examined if whole brain behaviors showed significant change across time with a repeated measures analysis of variance (ANOVA, p $<$= 0.05, FDR-corrected). Results show that the recruitment of the superior sensorimotor area (F = 3.8471, p = 0.0078) and anterior cingulate cortex (ACC, F = 2.5396, p = 0.0497) are significantly different in 6 months after stroke. Post-hoc pair t-tests show that compared with the baseline (the first time point), at 50-60 days and 140-150 days, respectively, since stroke recruitment of superior sensorimotor area (t = 2.9062, p = 0.0115) and ACC (t = 2.4940, p = 0.0257 0.0115) significantly increased. The integration of ACC (F = 2.5396, p = 0.0497) significantly differed in the 6-month post-stroke period. Post-hoc pair t-tests show that integration of ACC at 50-60 days since stroke is significantly reduced compared to baseline. Regarding flexibility, the ACC ( F = 4.0407, p = 0.0060) and left anterior insula ( F = 3.6657, p = 0.0101) significantly differed in the 6 months post-stroke period. Compared to baseline, the flexibility of ACC at 20-30, and 110-120 days since the stroke significantly increased  (t = 3.6422, p = 0.0026; t = 2.9459, p = 0.0106, post-hoc pair t-tests); the left anterior insula at days 160-200 and 80-120 since stroke significantly increased (t = 2.6366, p = 0.0195; t = 2.2100 p = 0.0442).

\subsection{Whole brain behaviors characterize good recovery and poor recovery groups.}

Given the significant difference in brain behavior compared with healthy control following stroke and the time effect on whole-brain behavior, we hypothesized that the good recovery and poor recovery groups would also have obvious alterations in terms of brain behaviors over the course of the 6-month post-stroke period. 

 The good recovery and poor recovery groups were assigned according to the patient's improvement of Brunnstrom recovery stage after six months. We first tracked the whole-brain average behavior in the two groups. Results show that since 80-90 days after the stroke, a significant between-group effect in recruitment and flexibility started to show. The good recovery group has significantly higher recruitment (t = 2.460, p = 0.028,FDR-corrected) but lower flexibility (t = -2.04, 0.031, FDR-corrected) than the poor recovery group. This significant alteration is continued till the sixth month and can be seen in day 110-120 days and 140-150 days since stroke (Recruitment, day 120: t = 2.414, p = 0.028; day 150: t =2.725, p = 0.0173; Flexibility, day 120: t = -0.250, p = 0.026, day 150: t = -3.21, p = 0.006. FDR-corrected). The integration did not observe significant changes between good and poor recovery groups at any time point. 

\begin{figure}[!t]
    \centering
    \includegraphics[width = 0.9\columnwidth]{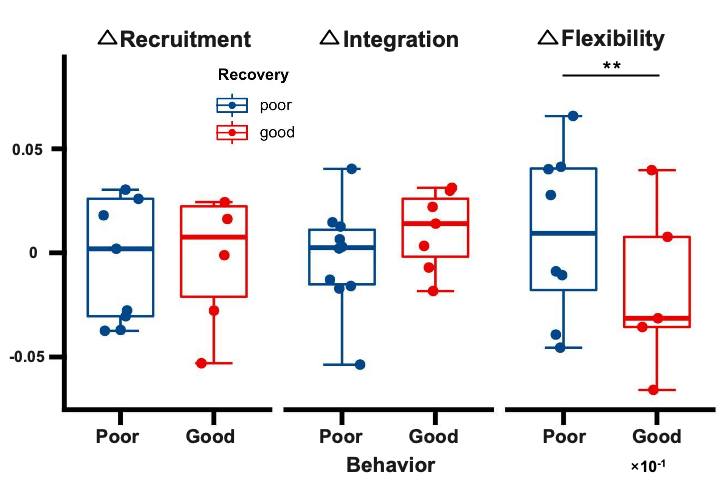}
    \caption{The recovery of global average behaviors between good and poor recovery groups.\label{fig: slope}}
\end{figure}
 
 In a subsequent step, whole brain recruitment and flexibility were probed if they were correlated with good or poor recovery that was measured by the Brunnstrom recovery score. Interestingly, the recruitment and flexibility can be only observed to have negative correlations with good and poor outcomes, respectively: the recruitment of the right posterior superior temporal gyrus (pSTG.R) and left bilateral frontal eye field (FEF.L) were significantly correlated with good outcomes (r = -0.0482, p = 0.0281; r = -0.4050, p = 0.0496, \autoref{fig: recovery}\textbf{B: Good panel}), whereas the flexibility of intraparietal sulcus (IPS) and left anterior insula were significantly correlated with poor outcome (r = -0.0482, p = 0.0281; r = -4050, p = 0.0496,\autoref{fig: recovery}\textbf{B: Poor panel}).

\begin{figure*}[!t]
    \centering
    \includegraphics[width = 0.80\textwidth]{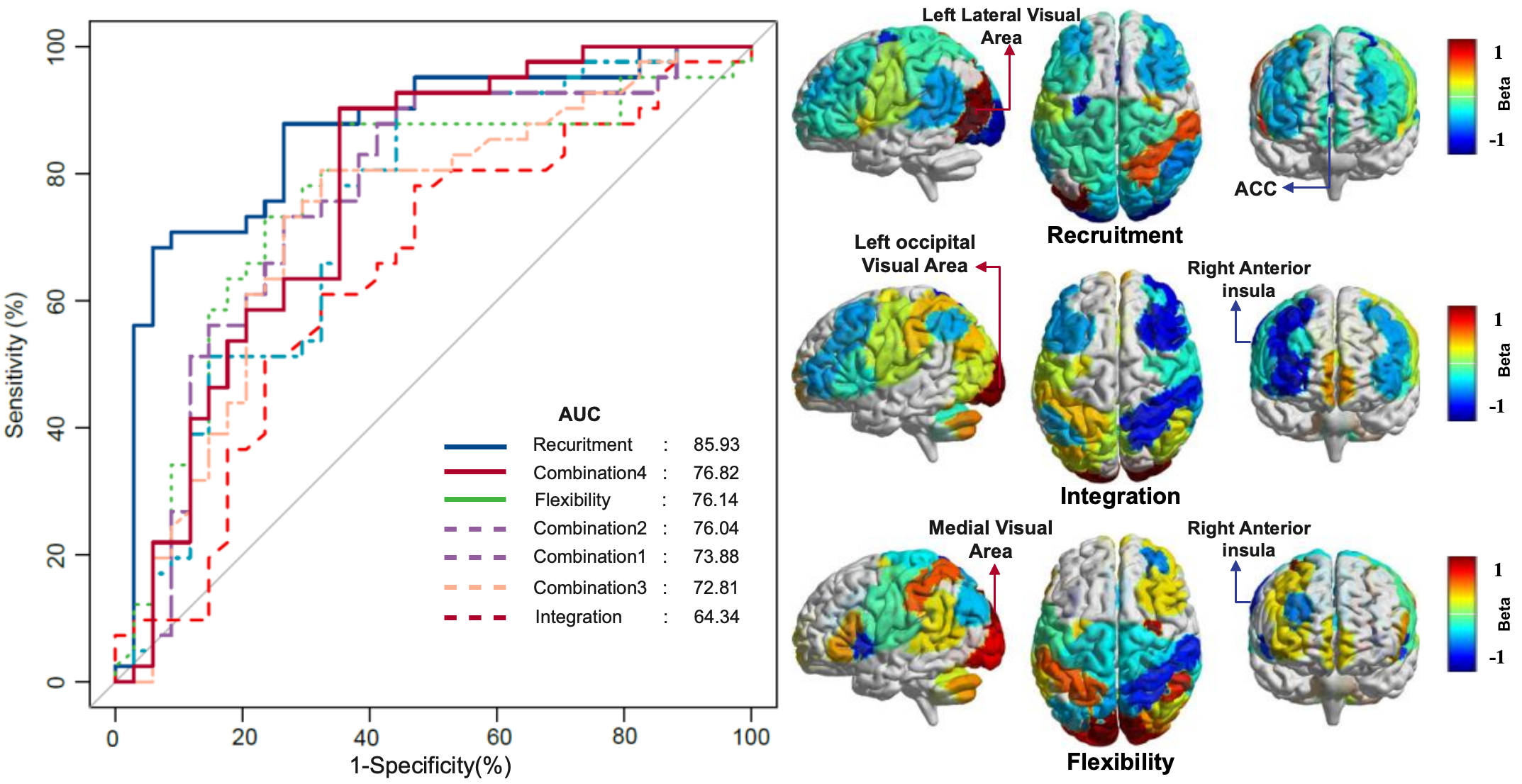}
    \caption{\textbf{(Left)} Utilize the whole-brain behavior to predict the degree of post-stroke motor function with ridge regression method.\textbf{(Right)} dynamic reconfiguration of whole brain area account for the contribution to motor function assessment.\label{fig: explanation}}
\end{figure*}

Stroke patients with motor impairment experience an extensive alteration in terms of dynamic behaviors, including significantly reduced recruitment, increased integration, and most-regions-increased flexibility. However, the patients with good recovery increased recruitment, and decreased flexibility, suggesting that recovery of dynamic behavior exhibits a positive correlation with motor rehabilitation. 
Next, the recovery of whole brain behavior across six months was then investigated. The slope of the recovery function served as the measure of whole-brain behavior recovery, which represented the direction and speed of recovery. In general, compared with the poor recovery group (\autoref{fig: slope}), the good recovery group exhibited slightly higher recovery in recruitment and integration as time went by. These recoveries are positively directed. Meanwhile, the good recovery group also observed a negative recovery in flexibility which was significantly lower than in patients with poor outcomes (p = .01). Diving into the direction of whole brain behaviors recovery, most brain regions in the good recovery group have opposite trends in terms of three whole brain behaviors. (The brain profiles can be seen in \autoref{fig: recovery} \textbf{C}). In terms of network, the good recovery group has decreased recruitment in DMN, FPN, LN, and CE and increased in VIS, SMN, and DAN. The VIS is the network with the most rapid growth in recruitment for the recovery group, which is also the only one that has the same trends as the poor recovery group. Others are all opposite to the poor recovery group. The network integration is all increased for good recovery groups except FPN, and the highest increasing trend is SMN. The poor group, in contrast, has decreased integration in DMN, VIS, DAN, and CE, and is opposite to the good recovery group. Regarding flexibility, good recovery seems to exhibit slow growth in VIS, SMN., and LN. Most networks were observed to have decreased. The poor recovery group has a relatively rapid increase in network flexibility, only the DMN shows a declined trend.

Collectively, good and poor recovery groups follow their divergent paths to recovery. This divergence starts 90 days after the stroke and manifests in higher recruitment and flexibility. Besides, while the behavior recovery can be only observed significantly differ in flexibility between good and poor recovery groups, the good recovery patients tend to have more positive recovery in recruitment and interaction.

\subsection{The pattern of whole brain behavior predicts the degree of post-stroke recovery}

Finally, we wished to establish a link between whole-brain measures and the degree of recovery and whether, from a clinical standpoint, brain behavior measures can improve the explanation of clinical outcomes.  

\begin{table*}[tbp]
\centering
    \fontsize{28pt}{28.5pt}\selectfont
    \caption{AUC, the global average $\beta$ and p-value for different prediction models on the various behaviors.\label{tab_ridge}}
    \renewcommand\arraystretch{1.5}
    \resizebox{\textwidth}{!}{
        \begin{tabular}{llllllllllllllllllllll}
            \toprule[3pt]

      & \multicolumn{3}{c}{Re}       & \multicolumn{3}{c}{In} & \multicolumn{3}{c}{Fe}       & \multicolumn{3}{c}{Re\_In} & \multicolumn{3}{c}{Re\_fe} & \multicolumn{3}{c}{In\_fe} & \multicolumn{3}{c}{Re\_in\_fe} \\
      & AUC & avg.$\beta$ & p-value                  & AUC  & avg.$\beta$  &p-value         & AUC & avg.$\beta$ & p-value                  & AUC      & avg.$\beta$      & p-value      & AUC      & avg.$\beta$      & p-value      & AUC      & avg.$\beta$      & p-value      & AUC        & avg.$\beta$       & p-value       \\

\midrule[3pt]
Lasso &  85.63   & 0.310  & \textless{}0.0001 & \textbf{64.24}    & 0.017  & 0.051  &   70.34  & 0.012  & \textless{}0.0001 & 70.18        &  -0.015 & \textless{}0.001        &     \textbf{76.84 }   &   -0.09     &   \textless{}0.001       & 70.81        &   -0.012      &    \textless{}0.001    &    70.32        & -0.19        &  \textless{}0.001   \\

Elastic Net &  84.90   & 0.120  & \textless{}0.001 & 58.34    & -0.250  & 0.121  &   63.14  & -0.007  & 0.069 & 71.21        &  0.035 & \textless{}0.001        &     73.57    &   -0.01     &   \textless{}0.001       & 69.11        &   0.007      &    \textless{}0.001    &    73.65        & -0.13        &  \textless{}0.001         \\

Ridge &  \textbf{85.93 }  & 0.090  & \textless{}0.0001 & 65.34    & -0.047  & 0.051  &   \textbf{76.14}  & 0.05  & \textless{}0.0001 & \textbf{73.88}        &  -0.085 & \textless{}0.001        &     76.04    &   -0.11     &   \textless{}0.001       & \textbf{72.81}        &   -0.047      &    \textless{}0.001    &    \textbf{76.82}        & -0.08        &  \textless{}0.001      \\
 \bottomrule[3pt]
        \end{tabular}
}
\end{table*}

There are 7 strategies implemented here which are the random but unique combinations for three behavior measurements. Combination 1 is recruitment and integration, 2 is the combination of recruitment and flexibility,  3 is the combination of integration and flexibility, and 4 is the combination of all three behaviors. Each strategy represents the possible network organization situation. The seven strategies will be fed into the ridge regression method to predict the stage of recovery(severe 0-3, mild 3-6). The results in \autoref{fig: explanation} present the prediction performance under these strategies. As can be seen, solely with the whole-brain recruitment, the RR model achieved the highest AUC of 85.93 (p$<$0.0001), which is a fair good performance compared to previous work\cite{wang2010dynamic,min2019power}. The worst performance is the RR model solely with whole-brain integration(AUC is 64.34). In the RR models with combination, the best one is the model with the three behaviors, achieving an AUC of 76.82, which is also the second-highest performance. This model is the closest one to the real brain decision system, as the whole brain regions would not have a single dynamic behavior according to the previous analysis. Of note, the RR model with recruitment and flexibility has a narrow gap to the top performance, which is 76.04. 

To determine the impact of prediction methods, two similar models: the lasso and elastic net, were implemented to test the prediction power of proposed dynamic behaviors. As can be seen in \autoref{tab_ridge}, the lasso and elastic net can achieve 85.63 and 84.90 with recruitment, respectively. Besides, the lasso model reached an AUC of 76.84 by inputting a combination of recruitment and flexibility, which is higher than that of the RR model.  The results demonstrate that dynamic behaviors in these methods still yield powerful predictive ability. However, unlike other machine learning methods, ridge regression can assign a weight to the dynamic behavior of each brain region by shrinking the coefficients. These weights indicate corresponding contributions in explaining the current recovery stage. The \autoref{fig: explanation} shows the project map of each brain's weight(the value of $\beta$). For the RR model with only recruitment, the left lateral visual area (unnormalized $\beta = 0.05$) contributes most to the prediction of recovery status, and ACC (unnormalized $\beta = -0.23$)is the biggest drag on performance. Recall the definition of recruitment.
the lost function of ACC to SMN might significantly interfere with or impede the recovery process.

\section{Discussion}

The network flexibility and function adaptability of the brain are supported by the brain function modularity\cite{RN24, RN27}. The behavior pattern of brain dynamic function modularity after stroke, therefore, presents how the brain responds to external changes in an adaptive and autonomous way. We here examined alterations of three whole-brain dynamic behaviors in 15 ischemic stroke patients and their relation to post-stroke motor function recovery. Three behaviors were identified, where the recruitment coefficient represents the allegiance of the brain region to its function network/system, the integration denotes the extent to which the regions are integrated into other new function networks, and the flexibility presents the speed of regions switching between networks. Those behaviors derived from the brain function module were proved to be altered extensively, which was represented by reduced recruitment, increased integration, and global ununified flexibility. Besides, the alterations of the behaviors after six months follow different paths to recovery and thus result in good and poor post-stroke outcomes. Most remarkably, the good recovery patients have positive directed recovery in recruitment and integration but negative flexibility. Those behaviors were further substantiated that they can be successfully applied in the prediction of post-stroke motor function recovery, where post-stroke recruitment has the best mapping performance.
    
\subsection{Alteration of the whole brain dynamic behaviors explains the function abnormalities and its recovery after stroke} 

Extensive differences in dynamic behaviors were detected in the 6 month period. Those significantly different regions spread over the whole brain area of various functional domains and vary across time. The dynamic behaviors were significantly correlated with the brunnstome recovery. In particular, the recruitment of the bilateral sensorimotor area positively correlated with the entire recovery score, indicating that the recovery of the sensorimotor area will promote motor rehabilitation. However, a frequently reported finding in previous studies is that abnormal brain behaviors have normalization trends\cite{RN18, RN4, RN31, RN24}, whereas the three dynamic behaviors identified in this paper did not exhibit obvious normalization towards the level observed in healthy control. The significant group effect still can be detected at the end of the follow-up. This is probably at the end of the follow-up, not all patients recover to normal. The heterogeneous groups with regard to stroke severity still exist. Nevertheless, the patients with good recovery in the patient groups have increased recruitment and decreased flexibility which reverses their original trends observed in stroke versus healthy controls, suggesting the recovery of recruitment and flexibility is parallel with function recovery. Besides, in the longitudinal study, the recruitment in the superior sensorimotor area and ACC, the integration and flexibility of ACC restored to normal level after an initial significant change when compared with the healthy controls. This finding suggested the normalization trend of abnormal behavior is not global (no significant longitudinal changes can be observed compared to the baseline in terms of the whole–brain average of three measures) but locally restrained to several regions. These regions are mainly resident in brain regions that reflect the dynamical reconfiguration within the brain network \cite{thompson2013neural} or in some communication hub \cite{van2011rich, RN20 }. On the other hand, the longitudinal follow-up is mostly at the chronic stage of post-stroke recovery. It has been reported that the acute or subacute stage is the high-speed recovery stage. The normalization trend should be more obvious if the earlier stages of patients are included. Hence, future studies need a long extension to cover the whole life of recovery to allow a thoroughly dynamic behavior analysis. 

\subsection{Dynamic behaviors in stroke and with post-stroke recovery} 

Increasingly recent work investigated dynamic behavior in post-stroke brain and its recovery resorting to dynamic functional connective changes(DFNC) analysis (the techniques detail and latest findings could see our previous review paper \cite{wu2022fmri}). With the temporal variability of FC in the fMRI scanning, Chen et al.\cite{RN9}, the FC variability slows down within the motor network after the brain, and Hu et al. \cite{RN21} substantiated that this reduction is at the acute stage and the brain modules the temporal variability to impact motor recovery from the acute stage to the early chronic stage. Others focused on the transient brain states derived from the time-varying function connectivity network. Bonkhoff et al. \cite{bonkhoff2020acute} presented four brain states of the motor network and showed that post-stroke brain altered the faction time of these states. Following that, in the global network investigation, they continued to prove that the brain would alter the dwell of transient brain states to promote realized recovery in stroke severity in the first 3 months after stroke ~\cite{RN19}. Similar work presented by Favaretto et al.~\cite{RN20}: they characterized the brain state with intact cortical-subcortical communication, and they found that the brain shifts between states to allow the recovery to be better explained. Beyond motor function,  Duncan et al.\cite{duncan2018changes} showed that brain states correlated with post-stroke aphasia severity, and Wang et al. \cite{wang2020imbalance} found the brain alters fraction times in patients with mid-brain lesions. 

Collectively, previous studies related to brain dynamic behaviors convey a point that the brain will dynamically readjust itself as much as possible to eliminate the stroke effects and, thus, toward recovery. In this work, based on the inherent function modularity of the brain, we confirm this finding and depict an image of how this adjustment happens by modeling autonomic brain behaviors. In addition, by tracking the time course of these brain behaviors' recovery, we determined the strength and direction of the brain adjustments. The relation between the recovery of behavior and function recovery score suggests proactive readjustments play a key role in post-stroke recovery. 

\subsection{Method considerations in the post-stroke recovery with dynamic behaviors} 

There are several method considerations in the post-stroke recovery with dynamic behaviors:

First, the process of dynamic recognition involves the determination of hyperparameters. The first is the sliding width and shifting step in DFNC estimation.   While there is a debate, the sliding width with a range from 16-30TR  (the 30s to 60s), and shifting step with 1 TR  are proved to have less noise \cite{Preti2017,hindriks2016can} and are wildly accepted in literature \cite{RN19, RN20, bonkhoff2020acute,wu2023tracking}. Hence in the experiment, we opt for 22 TR (44s) and 1 TR (2.2 s). Second,  the $\gamma$ and $\omega$in the multilayer functional modularity detection. Recent work\cite{yang2021measurement} has suggested $\gamma = [1.0-1.1]$  and $\omega = [1.7-3.0]$ has less impact on the test-retest reliability of integration and flexibility, while $\gamma$ = [1.05-1.25] and $\omega$ = [1.2-3.0] on recruitment. The $\gamma$ = 1.0 and $\omega$ = 1.2 is in our process of optimizing the multilayer modularity. Hence, the option is basically close to the optimal range. However, this could be seen as our limitation of this work, where further experiments may need to include different combinations of $\gamma$ and $\omega$ parameters to demonstrate the robustness. Third, the modeling of brain behaviors' recovery. In this work, following the work by Siegel et al. \cite{RN18}, the recovery of brain behavior in each patient was modeled using a log function. However, the log function is too simple to capture the temporal variation in dynamic brain behaviors. Hence, more nonlinear functions and long-term observation of brain behaviors should be involved in further experiments, so that the time course and time-dependence of these brain behaviors can be depicted.

Besides, most post-stroke studies focus on several motor-relevant brain regions ~\cite{min2019power, wang2010dynamic, bonkhoff2020acute}, like primary motor cortex(known as M1) or secondary motor area. Whereas increasing studies tend to involve the whole brain network rather than restrained to motor-relevant areas \cite{RN19, RN20,wu2023dynamic, RN21}. This is because (1) beyond the motor function, the whole brain models enable the other function impairments induced by stroke to be analyzed, thus building up a comprehensive post-stroke compensation mechanism, and (2) Whole brain models allow the interaction between various networks (e.g., cortico-subcortical coordination \cite{RN20, van2011rich}) to be detected and to be linked to the motor deficit and their recovery.  Since this paper involves the dynamic readjustment of various networks, to figure out the motor network reconfiguration after stroke, we opted for whole brain dynamic behaviors analysis. The changes in dynamic behaviors involve sensorimotor, auditory, cognitive control, and
default mode network domains. Hence the results also support recent works that focal lesions impact the intact connection of remote networks \cite{guggisberg2019brain,rivier2023prediction,allali2018brain}. In addition,  an application of such dynamic behavior in post-stroke recovery prediction suggests the role that the brain region plays in post-stroke recovery, which may provide cues in future clinical enacting rehabilitation strategies.

Finally, even though it is difficult
to accurately predict patients’ level of recovery and,
consequently, to assign them to the appropriate treatment\cite{nijland2013accuracy}, there is a lot of effort into the prediction of post-stroke outcomes. Rehme et al. ~\cite{rehme2015identifying} utilized resting-state connectivity to classify patients with hand motor deficits and nonimpaired patients with 82.6–87.6\% accuracy. Bonkhoff et al.\cite{bonkhoff2021dynamic} achieved an accuracy of 0.896 in motor recovery prediction with a random forest classifier by using the time spent in brain connectivity states. The Bayesian hierarchical models built by Bonkhoff et al. \cite{RN19} later also demonstrated that inputting NIHSS at admission and dwell times of brain state can achieve an increased explained variance to 62.1\%. There were also studies using the ridge regression method that are similar to the one adopted in this work. Rivier et al. \cite{rivier2023prediction}. achieved performance of $R^2 = 0.68 $ with brain connectivity measures combined with other clinical features. Instead of looking at post-stroke motor recovery, Favaretto et al. ~\cite{RN20}. proved that the dynamic FC features are more behaviorally relevant with NIHSS. Collectively, our work agrees with these studies that (I) the prediction performance usually does not rest with the adopted approaches, according to the performance of alternative methods. Besides, our work demonstrates that whole brain recruitment can achieve the highest results, which differs from another common finding that (ii)solely single input generally cannot achieve the best performance. This is probably because the recruitment comes with a dynamic function modularity which itself is a highly integrated measurement. Also, this finding certainly implies that adding other clinical features, like age, initial impairment, or lesion volume, into our model could facilitate the final performance. Most importantly, this work utilized the ridge regression model as we were motivated by the interest in identifying the discriminative power of dynamic brain behaviors, and, thus, to infer possible mechanisms of neural recovery. By analyzing the contribution to the prediction performance, our work suggested that SAN plays a key role in post-stroke recovery.

\section{Conclusion}

Based on the dynamic function modularity, this paper presents a detailed analysis of post-stroke dynamic behaviors. The findings demonstrate that the functional network underwent dynamic remodeling after stroke. During the subsequent recovery period, the brain's ongoing adjustments lead to varying expected outcomes among patients. Patients who experience good recovery exhibit significantly higher levels of recruitment and flexibility. The recovery of flexibility demonstrates a pronounced increase during the recovery process in patients with good recovery. Furthermore, in our investigation into predicting post-stroke recovery status, the results suggest that measuring the whole-brain recruitment within the brain network holds promise as a potential tool for assessing and predicting motor recovery after stroke. These findings contribute to an enhanced understanding of the dynamic properties of brain networks and offer fresh insights into the underlying mechanisms governing the reorganization and integration of the brain network throughout the recovery process following stroke.

\section*{Data and code availability statement}

All code, including the Matlab scripts for neuroimaging data processing and dynamic function modularity calculation, and the Jupyter, as well as R notebooks for statistical evaluations and visualizations, will be publicly available on the GitHub homepage after formal publication. The raw fMRI and clinical data are available from the corresponding author on request.

\section*{Reference}

\bibliographystyle{IEEEtran}
\bibliography{main}

\begin{thebibliography}{10}
\providecommand{\url}[1]{#1}
\csname url@samestyle\endcsname
\providecommand{\newblock}{\relax}
\providecommand{\bibinfo}[2]{#2}
\providecommand{\BIBentrySTDinterwordspacing}{\spaceskip=0pt\relax}
\providecommand{\BIBentryALTinterwordstretchfactor}{4}
\providecommand{\BIBentryALTinterwordspacing}{\spaceskip=\fontdimen2\font plus
\BIBentryALTinterwordstretchfactor\fontdimen3\font minus
  \fontdimen4\font\relax}
\providecommand{\BIBforeignlanguage}[2]{{%
\expandafter\ifx\csname l@#1\endcsname\relax
\typeout{** WARNING: IEEEtran.bst: No hyphenation pattern has been}%
\typeout{** loaded for the language `#1'. Using the pattern for}%
\typeout{** the default language instead.}%
\else
\language=\csname l@#1\endcsname
\fi
#2}}
\providecommand{\BIBdecl}{\relax}
\BIBdecl

\bibitem{RN1}
G.~L. R. o.~S. Collaborators, ``Global, regional, and country-specific lifetime
  risks of stroke, 1990 and 2016,'' \emph{New England Journal of Medicine},
  vol. 379, no.~25, pp. 2429--2437, 2018.

\bibitem{RN2}
M.~Corbetta, L.~Ramsey, A.~Callejas, A.~Baldassarre, C.~D. Hacker, J.~S.
  Siegel, S.~V. Astafiev, J.~Rengachary, K.~Zinn, and C.~E. Lang, ``Common
  behavioral clusters and subcortical anatomy in stroke,'' \emph{Neuron},
  vol.~85, no.~5, pp. 927--941, 2015.

\bibitem{RN3}
V.~L. Feigin, E.~Nichols, T.~Alam, M.~S. Bannick, E.~Beghi, N.~Blake, W.~J.
  Culpepper, E.~R. Dorsey, A.~Elbaz, and R.~G. Ellenbogen, ``Global, regional,
  and national burden of neurological disorders, 1990–2016: a systematic
  analysis for the global burden of disease study 2016,'' \emph{The Lancet
  Neurology}, vol.~18, no.~5, pp. 459--480, 2019.

\bibitem{RN4}
F.~Buma, G.~Kwakkel, and N.~Ramsey, ``Understanding upper limb recovery after
  stroke,'' \emph{Restorative neurology and neuroscience}, vol.~31, no.~6, pp.
  707--722, 2013.

\bibitem{RN6}
C.~Grefkes and G.~R. Fink, ``Connectivity-based approaches in stroke and
  recovery of function,'' \emph{The Lancet Neurology}, vol.~13, no.~2, pp.
  206--216, 2014.

\bibitem{RN5}
T.~H. Murphy and D.~Corbett, ``Plasticity during stroke recovery: from synapse
  to behaviour,'' \emph{Nature reviews neuroscience}, vol.~10, no.~12, pp.
  861--872, 2009.

\bibitem{RN34}
T.~A. Bolton, E.~Morgenroth, M.~G. Preti, and D.~Van De~Ville, ``Tapping into
  multi-faceted human behavior and psychopathology using fmri brain dynamics,''
  \emph{Trends in Neurosciences}, vol.~43, no.~9, pp. 667--680, 2020.

\bibitem{RN33}
N.~S. Ward, ``Restoring brain function after stroke—bridging the gap between
  animals and humans,'' \emph{Nature Reviews Neurology}, vol.~13, no.~4, pp.
  244--255, 2017.

\bibitem{RN9}
J.~Chen, D.~Sun, Y.~Shi, W.~Jin, Y.~Wang, Q.~Xi, and C.~Ren, ``Alterations of
  static functional connectivity and dynamic functional connectivity in motor
  execution regions after stroke,'' \emph{Neuroscience letters}, vol. 686, pp.
  112--121, 2018.

\bibitem{RN8}
J.~Du, W.~Yao, J.~Li, F.~Yang, J.~Hu, Q.~Xu, L.~Liu, Q.~Lv, R.~Liu, and R.~Ye,
  ``Motor network reorganization after repetitive transcranial magnetic
  stimulation in early stroke patients: a resting state fmri study,''
  \emph{Neurorehabilitation and neural repair}, vol.~36, no.~1, pp. 61--68,
  2022.

\bibitem{RN15}
C.-h. Park, W.~H. Chang, S.~H. Ohn, S.~T. Kim, O.~Y. Bang, A.~Pascual-Leone,
  and Y.-H. Kim, ``Longitudinal changes of resting-state functional
  connectivity during motor recovery after stroke,'' \emph{Stroke}, vol.~42,
  no.~5, pp. 1357--1362, 2011.

\bibitem{RN32}
M.~P. Van~Meer, K.~Van Der~Marel, K.~Wang, W.~M. Otte, S.~El~Bouazati, T.~A.
  Roeling, M.~A. Viergever, J.~W.~B. van~der Sprenkel, and R.~M. Dijkhuizen,
  ``Recovery of sensorimotor function after experimental stroke correlates with
  restoration of resting-state interhemispheric functional connectivity,''
  \emph{Journal of Neuroscience}, vol.~30, no.~11, pp. 3964--3972, 2010.

\bibitem{RN31}
A.-M. Golestani, S.~Tymchuk, A.~Demchuk, B.~G. Goodyear, and V.-.~S. Group,
  ``Longitudinal evaluation of resting-state fmri after acute stroke with
  hemiparesis,'' \emph{Neurorehabilitation and neural repair}, vol.~27, no.~2,
  pp. 153--163, 2013.

\bibitem{RN17}
H.~E. Yang, S.~Kyeong, S.~H. Lee, W.-J. Lee, S.~W. Ha, S.~M. Kim, H.~Kang,
  W.~M. Lee, C.~S. Kang, and D.~H. Kim, ``Structural and functional
  improvements due to robot-assisted gait training in the stroke-injured
  brain,'' \emph{Neuroscience letters}, vol. 637, pp. 114--119, 2017.

\bibitem{RN16}
S.~I. Ktena, M.~D. Schirmer, M.~R. Etherton, A.-K. Giese, C.~Tuozzo, B.~B.
  Mills, D.~Rueckert, O.~Wu, and N.~S. Rost, ``Brain connectivity measures
  improve modeling of functional outcome after acute ischemic stroke,''
  \emph{Stroke}, vol.~50, no.~10, pp. 2761--2767, 2019.

\bibitem{RN18}
J.~S. Siegel, B.~A. Seitzman, L.~E. Ramsey, M.~Ortega, E.~M. Gordon, N.~U.
  Dosenbach, S.~E. Petersen, G.~L. Shulman, and M.~Corbetta, ``Re-emergence of
  modular brain networks in stroke recovery,'' \emph{Cortex}, vol. 101, pp.
  44--59, 2018.

\bibitem{RN35}
B.~Mišić and O.~Sporns, ``From regions to connections and networks: new
  bridges between brain and behavior,'' \emph{Current opinion in neurobiology},
  vol.~40, pp. 1--7, 2016.

\bibitem{RN38}
M.~Pekna, M.~Pekny, and M.~Nilsson, ``Modulation of neural plasticity as a
  basis for stroke rehabilitation,'' \emph{Stroke}, vol.~43, no.~10, pp.
  2819--2828, 2012.

\bibitem{RN37}
K.~Wu, B.~Jelfs, K.~Neville, and J.~Q. Fang, ``fmri-based static and dynamic
  functional connectivity analysis for post-stroke motor dysfunction patient: A
  review,'' \emph{arXiv preprint arXiv:2301.07171}, 2022.

\bibitem{RN36}
C.~Wang, W.~Qin, J.~Zhang, T.~Tian, Y.~Li, L.~Meng, X.~Zhang, and C.~Yu,
  ``Altered functional organization within and between resting-state networks
  in chronic subcortical infarction,'' \emph{Journal of Cerebral Blood Flow \&
  Metabolism}, vol.~34, no.~4, pp. 597--605, 2014.

\bibitem{RN22}
E.~A. Allen, E.~Damaraju, S.~M. Plis, E.~B. Erhardt, T.~Eichele, and V.~D.
  Calhoun, ``Tracking whole-brain connectivity dynamics in the resting state,''
  \emph{Cerebral cortex}, vol.~24, no.~3, pp. 663--676, 2014.

\bibitem{RN23}
V.~D. Calhoun, R.~Miller, G.~Pearlson, and T.~Adalı, ``The chronnectome:
  time-varying connectivity networks as the next frontier in fmri data
  discovery,'' \emph{Neuron}, vol.~84, no.~2, pp. 262--274, 2014.

\bibitem{RN24}
O.~Sporns, ``Network attributes for segregation and integration in the human
  brain,'' \emph{Current opinion in neurobiology}, vol.~23, no.~2, pp.
  162--171, 2013.

\bibitem{RN28}
N.~Kashtan and U.~Alon, ``Spontaneous evolution of modularity and network
  motifs,'' \emph{Proceedings of the National Academy of Sciences}, vol. 102,
  no.~39, pp. 13\,773--13\,778, 2005.

\bibitem{RN30}
J.~R. Cohen and M.~D'Esposito, ``The segregation and integration of distinct
  brain networks and their relationship to cognition,'' \emph{Journal of
  Neuroscience}, vol.~36, no.~48, pp. 12\,083--12\,094, 2016.

\bibitem{RN29}
H.~Mohr, U.~Wolfensteller, R.~F. Betzel, B.~Mišić, O.~Sporns, J.~Richiardi,
  and H.~Ruge, ``Integration and segregation of large-scale brain networks
  during short-term task automatization,'' \emph{Nature communications},
  vol.~7, no.~1, p. 13217, 2016.

\bibitem{RN27}
K.~Finc, K.~Bonna, X.~He, D.~M. Lydon-Staley, S.~Kühn, W.~Duch, and D.~S.
  Bassett, ``Dynamic reconfiguration of functional brain networks during
  working memory training,'' \emph{Nature communications}, vol.~11, no.~1, p.
  2435, 2020.

\bibitem{RN39}
G.~Gifford, N.~Crossley, M.~J. Kempton, S.~Morgan, P.~Dazzan, J.~Young, and
  P.~McGuire, ``Resting state fmri based multilayer network configuration in
  patients with schizophrenia,'' \emph{NeuroImage: Clinical}, vol.~25, p.
  102169, 2020.

\bibitem{RN40}
X.~He, D.~S. Bassett, G.~Chaitanya, M.~R. Sperling, L.~Kozlowski, and J.~I.
  Tracy, ``Disrupted dynamic network reconfiguration of the language system in
  temporal lobe epilepsy,'' \emph{Brain}, vol. 141, no.~5, pp. 1375--1389,
  2018.

\bibitem{RN41}
S.~Han, Q.~Cui, X.~Wang, L.~Li, D.~Li, Z.~He, X.~Guo, Y.~Fan, J.~Guo, and
  W.~Sheng, ``Resting state functional network switching rate is differently
  altered in bipolar disorder and major depressive disorder,'' \emph{Human
  brain mapping}, vol.~41, no.~12, pp. 3295--3304, 2020.

\bibitem{RN19}
A.~K. Bonkhoff, M.~D. Schirmer, M.~Bretzner, M.~Etherton, K.~Donahue,
  C.~Tuozzo, M.~Nardin, A.~Giese, O.~Wu, and V.~D.~Calhoun, ``Abnormal dynamic
  functional connectivity is linked to recovery after acute ischemic stroke,''
  \emph{Human brain mapping}, vol.~42, no.~7, pp. 2278--2291, 2021.

\bibitem{RN20}
C.~Favaretto, M.~Allegra, G.~Deco, N.~V. Metcalf, J.~C. Griffis, G.~L. Shulman,
  A.~Brovelli, and M.~Corbetta, ``Subcortical-cortical dynamical states of the
  human brain and their breakdown in stroke,'' \emph{Nature communications},
  vol.~13, no.~1, p. 5069, 2022.

\bibitem{RN21}
J.~Hu, J.~Du, Q.~Xu, F.~Yang, F.~Zeng, Y.~Weng, X.-j. Dai, R.~Qi, X.~Liu, and
  G.~Lu, ``Dynamic network analysis reveals altered temporal variability in
  brain regions after stroke: a longitudinal resting-state fmri study,''
  \emph{Neural plasticity}, vol. 2018, 2018.

\bibitem{shah1986stroke}
S.~Shah, S.~Harasymiw, and P.~Stahl, ``Stroke rehabilitation: outcome based on
  brunnstrom recovery stages,'' \emph{The Occupational Therapy Journal of
  Research}, vol.~6, no.~6, pp. 365--376, 1986.

\bibitem{RNre}
D.~S. Bassett, M.~Yang, N.~F. Wymbs, and S.~T. Grafton, ``Learning-induced
  autonomy of sensorimotor systems,'' \emph{Nature Neuroscience}, vol.~18,
  no.~5, pp. 744--751, 2015.

\bibitem{RNfe}
J.~S. Siegel, L.~E. Ramsey, A.~Z. Snyder, N.~V. Metcalf, R.~V. Chacko,
  K.~Weinberger, A.~Baldassarre, C.~D. Hacker, G.~L. Shulman, and M.~Corbetta,
  ``Disruptions of network connectivity predict impairment in multiple
  behavioral domains after stroke,'' \emph{Proceedings of the National Academy
  of Sciences}, vol. 113, no.~30, pp. E4367--E4376, 2016.

\bibitem{wang2010dynamic}
L.~Wang, C.~Yu, H.~Chen, W.~Qin, Y.~He, F.~Fan, Y.~Zhang, M.~Wang, K.~Li,
  Y.~Zang \emph{et~al.}, ``Dynamic functional reorganization of the motor
  execution network after stroke,'' \emph{Brain}, vol. 133, no.~4, pp.
  1224--1238, 2010.

\bibitem{min2019power}
Y.-S. Min, J.~W. Park, K.~E. Jang, H.~J. Lee, J.~Lee, Y.-S. Lee, T.-D. Jung,
  and Y.~Chang, ``Power spectral density analysis of long-term motor recovery
  in patients with subacute stroke,'' \emph{Neurorehabilitation and Neural
  Repair}, vol.~33, no.~1, pp. 38--46, 2019.

\bibitem{thompson2013neural}
G.~J. Thompson, M.~D. Merritt, W.-J. Pan, M.~E. Magnuson, J.~K. Grooms,
  D.~Jaeger, and S.~D. Keilholz, ``Neural correlates of time-varying functional
  connectivity in the rat,'' \emph{Neuroimage}, vol.~83, pp. 826--836, 2013.

\bibitem{van2011rich}
M.~P. Van Den~Heuvel and O.~Sporns, ``Rich-club organization of the human
  connectome,'' \emph{Journal of Neuroscience}, vol.~31, no.~44, pp.
  15\,775--15\,786, 2011.

\bibitem{wu2022fmri}
K.~Wu, B.~Jelfs, K.~Neville, and J.~Q. Fang, ``fmri-based static and dynamic
  functional connectivity analysis for post-stroke motor dysfunction patient: A
  review,'' \emph{arXiv preprint arXiv:2301.07171}, 2022.

\bibitem{bonkhoff2020acute}
A.~K. Bonkhoff, F.~A. Espinoza, H.~Gazula, V.~M. Vergara, L.~Hensel,
  J.~Michely, T.~Paul, A.~K. Rehme, L.~J. Volz, G.~R. Fink \emph{et~al.},
  ``Acute ischaemic stroke alters the brain’s preference for distinct dynamic
  connectivity states,'' \emph{Brain}, vol. 143, no.~5, pp. 1525--1540, 2020.

\bibitem{duncan2018changes}
E.~S. Duncan and S.~L. Small, ``Changes in dynamic resting state network
  connectivity following aphasia therapy,'' \emph{Brain imaging and behavior},
  vol.~12, pp. 1141--1149, 2018.

\bibitem{wang2020imbalance}
Y.~Wang, C.~Wang, P.~Miao, J.~Liu, Y.~Wei, L.~Wu, K.~Wang, and J.~Cheng, ``An
  imbalance between functional segregation and integration in patients with
  pontine stroke: a dynamic functional network connectivity study,''
  \emph{NeuroImage: Clinical}, vol.~28, p. 102507, 2020.

\bibitem{Preti2017}
M.~G. Preti, T.~A. Bolton, and D.~Van De~Ville, ``The dynamic functional
  connectome: State-of-the-art and perspectives,'' \emph{Neuroimage}, vol. 160,
  pp. 41--54, 2017.

\bibitem{hindriks2016can}
R.~Hindriks, M.~H. Adhikari, Y.~Murayama, M.~Ganzetti, D.~Mantini, N.~K.
  Logothetis, and G.~Deco, ``Can sliding-window correlations reveal dynamic
  functional connectivity in resting-state fmri?'' \emph{Neuroimage}, vol. 127,
  pp. 242--256, 2016.

\bibitem{wu2023tracking}
K.~Wu, B.~Jelfs, S.~S. Mahmoud, K.~Neville, and J.~Q. Fang, ``Tracking
  functional network connectivity dynamics in the elderly,'' \emph{Frontiers in
  Neuroscience}, vol.~17, p. 1146264, 2023.

\bibitem{yang2021measurement}
Z.~Yang, Q.~K. Telesford, A.~R. Franco, R.~Lim, S.~Gu, T.~Xu, L.~Ai, F.~X.
  Castellanos, C.-G. Yan, S.~Colcombe \emph{et~al.}, ``Measurement reliability
  for individual differences in multilayer network dynamics: Cautions and
  considerations,'' \emph{NeuroImage}, vol. 225, p. 117489, 2021.

\bibitem{wu2023dynamic}
K.~Wu, B.~Jelfs, K.~Neville, W.~He, and Q.~Fang, ``Dynamic reconfiguration of
  brain functional network in stroke,'' 2023.

\bibitem{guggisberg2019brain}
A.~G. Guggisberg, P.~J. Koch, F.~C. Hummel, and C.~M. Buetefisch, ``Brain
  networks and their relevance for stroke rehabilitation,'' \emph{Clinical
  Neurophysiology}, vol. 130, no.~7, pp. 1098--1124, 2019.

\bibitem{rivier2023prediction}
C.~Rivier, M.~G. Preti, P.~Nicolo, D.~Van De~Ville, A.~G. Guggisberg, and
  E.~Pirondini, ``Prediction of post-stroke motor recovery benefits from
  measures of sub-acute widespread network damages,'' \emph{Brain
  Communications}, vol.~5, no.~2, p. fcad055, 2023.

\bibitem{allali2018brain}
G.~Allali, H.~M. Blumen, H.~Devanne, E.~Pirondini, A.~Delval, and D.~Van
  De~Ville, ``Brain imaging of locomotion in neurological conditions,''
  \emph{Neurophysiologie Clinique}, vol.~48, no.~6, pp. 337--359, 2018.

\bibitem{nijland2013accuracy}
R.~H. Nijland, E.~E. Van~Wegen, B.~C. Harmeling-van~der Wel, G.~Kwakkel, and
  E.~P. of~Functional Outcome After Stroke (EPOS)~Investigators, ``Accuracy of
  physical therapists' early predictions of upper-limb function in hospital
  stroke units: the epos study,'' \emph{Physical therapy}, vol.~93, no.~4, pp.
  460--469, 2013.

\bibitem{rehme2015identifying}
A.~K. Rehme, L.~J. Volz, D.-L. Feis, I.~Bomilcar-Focke, T.~Liebig, S.~B.
  Eickhoff, G.~R. Fink, and C.~Grefkes, ``Identifying neuroimaging markers of
  motor disability in acute stroke by machine learning techniques,''
  \emph{Cerebral cortex}, vol.~25, no.~9, pp. 3046--3056, 2015.

\bibitem{bonkhoff2021dynamic}
A.~K. Bonkhoff, A.~K. Rehme, L.~Hensel, C.~Tscherpel, L.~J. Volz, F.~A.
  Espinoza, H.~Gazula, V.~M. Vergara, G.~R. Fink, V.~D. Calhoun \emph{et~al.},
  ``Dynamic connectivity predicts acute motor impairment and recovery
  post-stroke,'' \emph{Brain communications}, vol.~3, no.~4, p. fcab227, 2021.

\end{thebibliography}




\begin{table*}[tbp]
\centering
    \fontsize{28pt}{28.5pt}\selectfont
    \caption{Differences in whole brain recruitment between controls and stroke patients at five-time points after stroke.\label{tab1}}
    \renewcommand\arraystretch{1.5}
    \resizebox{\textwidth}{!}{
        \begin{tabular}{lcclcclcclcclcc}
            \toprule[3pt]
    
\multicolumn{3}{l}{1st versus controls}                       & \multicolumn{3}{l}{2nd versus controls}                       & \multicolumn{3}{l}{3rd versus controls}                       & \multicolumn{3}{l}{4th versus controls}                       & \multicolumn{3}{l}{5th versus controls}                       \\

Brain region                       & T-value       & p-value          & Brain region                       & T-value       & p-value      & Brain region                       & T-value       & p-value      & Brain region                       & T-value       & p-value         & Brain region                       & T-value       & p-value        \\

\midrule[3pt]
      DAN.FEF(L) & -2.774 & $<$0.001 & DAN.IPS(R) & -2.626 & 0.034 & DAN.IPS(R) & -2.353 & 0.048 & DAN.FEF(L) & -2.602 & 0.034 & LN.IFG(R) & -2.415 & 0.048 \\
DAN.IPS(L) & -2.275 & 0.03 & FPN.PPC(L) & -2.515 & 0.038 & FPN.PPC(L) & -2.466 & 0.04 & DAN.FEF(R) & -3.062 & 0.013 & SAN.ACC & -2.607 & 0.035 \\
DAN.IPS(R) & -2.115 & 0.043 & LN.IFG(L) & -2.664 & 0.033 & LN.IFG(R) & -2.749 & 0.022 & DAN.IPS(L) & -2.867 & 0.019 & SAN.AInsula(L) & -4.073 & 0.002 \\
LN.IFG(L) & -3.542 & 0.001 & LN.IFG(R) & -2.727 & 0.031 & SAN.ACC & -3.443 & 0.004 & DAN.IPS(R) & -3.573 & 0.005 & SAN.AInsula(R) & -3.960 & 0.002 \\
LN.IFG(R) & -3.358 & 0.002 & SAN.ACC & -4.645 & $<$0.001 & SAN.AInsula(L) & -5.066 & $<$0.001 & SAN.ACC & -5.670 & $<$0.001 & SAN.RPFC(L) & -3.416 & 0.008 \\
SAN.ACC & -2.963 & 0.06 & SAN.AInsula(L) & -6.981 & $<$0.001 & SAN.AInsula(R) & -4.483 & $<$0.001 & SAN.AInsula(L) & -5.253 & $<$0.001 & SAN.RPFC(R) & -3.235 & 0.009 \\
SAN.AInsula(L) & -4.567 & $<$0.001 & SAN.AInsula(R) & -5.270 & $<$0.001 & SAN.RPFC(L) & -4.263 & $<$0.001 & SAN.AInsula(R) & -5.048 & $<$0.001 & SAN.SMG(L) & -2.870 & 0.02 \\
SAN.AInsula(R) & -4.528 & $<$0.001 & SAN.RPFC(L) & -4.737 & $<$0.001 & SAN.RPFC(R) & -3.440 & 0.004 & SAN.RPFC(L) & -4.561 & $<$0.001 & SAN.SMG(R) & -2.445 & 0.048 \\
SAN.RPFC(L) & -3.698 & $<$0.001 & SAN.RPFC(R) & -3.984 & 0.001 & SAN.SMG(L) & -4.333 & $<$0.001 & SAN.RPFC(R) & -3.520 & 0.005 & SMN.Lateral(L) & -3.637 & 0.005 \\
SAN.RPFC(R) & -3.119 & $<$0.001 & SAN.SMG(L) & -5.048 & $<$0.001 & SAN.SMG(R) & -3.520 & 0.004 & SAN.SMG(L) & -4.158 & 0.001 & SMN.Lateral(R) & -3.358 & 0.009 \\
SAN.SMG(L) & -3.135 & $<$0.001 & SAN.SMG(R) & -3.549 & 0.004 & SMN.Lateral(L) & -3.957 & 0.001 & SAN.SMG(R) & -3.357 & 0.007 & SMN.Superior & -2.949 & 0.018 \\
SAN.SMG(R) & -2.936 & $<$0.001 & SMN.Lateral(L) & -2.462 & 0.04 & SMN.Lateral(R) & -3.943 & 0.001 & SMN.Lateral(L) & -2.219 & 0.069 & VIS.Lateral(L) & -6.930 & $<$0.001 \\
SMN.Lateral(L) & -2.783 & $<$0.001 & SMN.Lateral(R) & -2.409 & 0.042 & SMN.Superior & -3.017 & 0.012 & VIS.Lateral(L) & -4.492 & $<$0.001 & VIS.Lateral(R) & -5.921 & $<$0.001 \\
SMN.Lateral(R) & -3.049 & $<$0.001 & VIS.Lateral(L) & -5.065 & $<$0.001 & VIS.Lateral(L) & -6.911 & $<$0.001 & VIS.Lateral(R) & -4.981 & $<$0.001 & VIS.Medial & -4.419 & 0.001 \\
SMN.Superior & -2.469 & 0.01 & VIS.Lateral(R) & -4.699 & $<$0.001 & VIS.Lateral(R) & -7.704 & $<$0.001 & VIS.Medial & -2.987 & 0.0154 & VIS.Occipital & -3.293 & 0.009 \\
VIS.Lateral(L) & -6.982 & $<$0.001 & VIS.Medial & -3.584 & 0.004 & VIS.Medial & -5.189 & $<$0.001 &  &  &  &  &  &  \\
VIS.Lateral(R) & -5.897 & $<$0.001 & VIS.Occipital & -2.530 & 0.038 & VIS.Occipital & -4.258 & $<$0.001 &  &  &  &  &  &  \\
VIS.Medial & -4.516 & $<$0.001 &  &  &  &  &  &  &  &  &  &  &  &  \\
VIS.Occipital & -4.957 & $<$0.001 &  &  &  &  &  &  &  &  &  &  &  &  \\
 \bottomrule[3pt]
        \end{tabular}
}
\end{table*}

\begin{table*}[tbp]
\centering
    \fontsize{28pt}{28.5pt}\selectfont
    \caption{Differences in whole brain integration between controls and stroke patients at five-time points after stroke.\label{tab1}}
    \renewcommand\arraystretch{1.5}
    \resizebox{\textwidth}{!}{
        \begin{tabular}{lcclcclcclcclcc}
            \toprule[3pt]
    
\multicolumn{3}{l}{1st versus controls}                       & \multicolumn{3}{l}{2nd versus controls}                       & \multicolumn{3}{l}{3rd versus controls}                       & \multicolumn{3}{l}{4th versus controls}                       & \multicolumn{3}{l}{5th versus controls}                       \\

Brain region                       & T-value       & p-value          & Brain region                       & T-value       & p-value      & Brain region                       & T-value       & p-value      & Brain region                       & T-value       & p-value         & Brain region                       & T-value       & p-value        \\

\midrule[3pt]
VIS.Lateral(L) & 3.967 & 0.005 & SAN.AInsula(L) & 3.383 & 0.030 & FPN.LPFC(R) & 2.796 & 0.049 & FPN.PPC(L) & 2.401 & 0.050 & VIS.Lateral(L) & 3.573 & 0.021 \\
VIS.Lateral(R) & 4.001 & 0.005 & SAN.AInsula(R) & 2.935 & 0.045 & FPN.PPC(L) & 3.271 & 0.023 & FPN.PPC(R) & 2.539 & 0.040 & VIS.Lateral(R) & 4.037 & 0.012 \\
VIS.Medial & 3.734 & 0.007 & SAN.RPFC(L) & 2.851 & 0.045 & SAN.SMG(L) & 3.384 & 0.023 & LN.IFG(L) & 3.360 & 0.009 & VIS.Medial & 3.130 & 0.043 \\
VIS.Occipital & 6.322 & 0.000 & VIS.Lateral(L) & 3.271 & 0.030 & SAN.SMG(R) & 2.863 & 0.049 & LN.IFG(R) & 2.527 & 0.040 &  &  &  \\
 &  &  & VIS.Lateral(R) & 3.770 & 0.025 & VIS.Lateral(L) & 3.957 & 0.008 & SAN.ACC & 3.635 & 0.005 &  &  &  \\
 &  &  & VIS.Medial & 2.766 & 0.045 & VIS.Lateral(R) & 4.699 & 0.002 & SAN.AInsula(L) & 3.823 & 0.004 &  &  &  \\
 &  &  & VIS.Occipital & 2.774 & 0.045 &  &  &  & SAN.AInsula(R) & 4.271 & 0.001 &  &  &  \\
 &  &  &  &  &  &  &  &  & SAN.RPFC(L) & 3.079 & 0.014 &  &  &  \\
 &  &  &  &  &  &  &  &  & SAN.RPFC(R) & 2.962 & 0.016 &  &  &  \\
 &  &  &  &  &  &  &  &  & SAN.SMG(L) & 3.268 & 0.010 &  &  &  \\
 &  &  &  &  &  &  &  &  & SAN.SMG(R) & 3.059 & 0.014 &  &  &  \\
 &  &  &  &  &  &  &  &  & VIS.Lateral(L) & 4.245 & 0.001 &  &  &  \\
 &  &  &  &  &  &  &  &  & VIS.Lateral(R) & 5.265 & 0.000 &  &  &  \\
 &  &  &  &  &  &  &  &  & VIS.Medial & 4.447 & 0.001 &  &  &  \\
 &  &  &  &  &  &  &  &  & VIS.Occipital & 5.190 & 0.000 &  &  &  \\
 
 \bottomrule[3pt]
        \end{tabular}
}
\end{table*}

\begin{table*}[tbp]
\centering
    \fontsize{28pt}{28.5pt}\selectfont
    \caption{Differences in whole brain flexibility between controls and stroke patients at five-time points after stroke.\label{tab1}}
    \renewcommand\arraystretch{1.5}
    \resizebox{\textwidth}{!}{
        \begin{tabular}{lcclcclcclcclcc}
            \toprule[3pt]
    
\multicolumn{3}{l}{1st versus controls}                       & \multicolumn{3}{l}{2nd versus controls}                       & \multicolumn{3}{l}{3rd versus controls}                       & \multicolumn{3}{l}{4th versus controls}                       & \multicolumn{3}{l}{5th versus controls}                       \\

Brain region                       & T-value       & p-value          & Brain region                       & T-value       & p-value      & Brain region                       & T-value       & p-value      & Brain region                       & T-value       & p-value         & Brain region                       & T-value       & p-value        \\

\midrule[3pt]

CE.Anterior & -2.291 & 0.030 & SAN.ACC & 2.871 & 0.008 & SAN.AInsula(L) & 2.073 & 0.047 & CE.Anterior & -2.148 & 0.041 &  &  &  \\
DMN.PCC & -2.393 & 0.024 & SAN.SMG(L) & 2.197 & 0.036 & SAN.AInsula(R) & 2.824 & 0.009 & SAN.ACC & 2.991 & 0.006 &  &  &  \\
VIS.Lateral(L) & 2.132 & 0.042 &  &  &  & SAN.RPFC(R) & 3.081 & 0.005 &  &  &  &  &  &  \\
 &  &  &  &  &  & SMN.Lateral(L) & 3.169 & 0.004 &  &  &  &  &  &  \\
 &  &  &  &  &  & VIS.Lateral(L) & 2.197 & 0.036 &  &  &  &  &  &  \\
 &  &  &  &  &  & VIS.Lateral(R) & 2.108 & 0.044 &  &  &  &  &  &  \\
 &  &  &  &  &  & VIS.Occipital & 2.413 & 0.023 &  &  &  &  &  &  \\

 \bottomrule[3pt]
        \end{tabular}
}

\end{table*}
\end{document}